\documentclass[final,3p,times,twocolumn,authoryear]{elsarticle}

\usepackage{amssymb}

\def\simlt{\mathrel{\rlap{\lower 3pt\hbox{$\sim$}}\raise 2.0pt\hbox{$<$}}}
\def\simgt{\mathrel{\rlap{\lower 3pt\hbox{$\sim$}} \raise 2.0pt\hbox{$>$}}}
\def\lsim{\mathrel{\rlap{\lower 3pt\hbox{$\sim$}}\raise 2.0pt\hbox{$<$}}}
\def\gsim{\mathrel{\rlap{\lower 3pt\hbox{$\sim$}} \raise  2.0pt\hbox{$>$}}}

\def\apjs{ApJS}

\journal{Journal of High Energy Astrophysics}

\begin{document}

\begin{frontmatter}



\title{Lorentz factor - Beaming Corrected Energy/Luminosity Correlations and GRB Central Engine Models}


\author[label1,label2,label3]{Shuang-Xi Yi\corref{dip}}
\ead{yishuangxi@nju.edu.cn}
\author[label4]{Wei-Hua Lei\corref{dip}}
\ead{leiwh@hust.edu.cn}
\author[label3]{Bing Zhang}
\author[label2]{Zi-Gao Dai}
\author[label5]{Xue-Feng Wu}
\author[label6,label7]{En-Wei Liang}
\address[label1]{College of Physics and Engineering, Qufu Normal University, Qufu 273165, China}
\def\astrobj#1{#1}
\address[label2]{School of Astronomy and Space Science, Nanjing University, Nanjing 210093, China}
\address[label3]{Department of Physics and Astronomy, University of Nevada Las Vegas, NV 89154, USA}
\address[label4]{School of Physics, Huazhong University of Science and Technology, Wuhan 430074, China}
\address[label5]{Purple Mountain Observatory, Chinese Academy of Sciences, Nanjing 210008, China}
\address[label6]{GXU-NAOC Center for Astrophysics and Space Sciences, Department of Physics, Guangxi University, Nanning 530004}
\address[label7]{Guangxi Key Laboratory for the Relativistic Astrophysics, Nanning 530004, China}
\cortext[dip]{Corresponding author.}

\begin{abstract}
We work on a GRB sample whose initial Lorentz factors ($\Gamma_0$) are constrained with the afterglow onset method and the jet opening angles ($\theta_{\rm j}$) are determined by the jet break time. We confirm the $\Gamma_0$ - $E_{\gamma,\rm iso}$ correlation by Liang et al. (2010), and the $\Gamma_0$ - $L_{\gamma,\rm iso}$ correlation by L{\"u} et al. (2012). Furthermore, we find correlations between $\Gamma_0$ and the beaming corrected $\gamma$-ray energy ($E_{\gamma}$) and mean $\gamma$-ray luminosity ($L_{\gamma}$). By also including the kinetic energy of the afterglow,
we find rough correlations (with larger scatter) between $\Gamma_0$ and the total ($\gamma$-ray plus kinetic) energy and the total mean luminosity, both for isotropic values and beaming corrected values: these correlations allow us to test the data with GRB central engine models.
Limiting our sample to the GRBs that likely have a black hole central engine, we compare the data with theoretical predictions of two types of jet launching mechanisms from BHs, i.e. the non-magnetized $\nu \bar\nu$ - annihilation mechanism, and the strongly magnetized Blandford-Znajek (BZ) mechanism. We find that the data are more consistent with the latter mechanism, and discuss the implications of our findings for GRB jet composition.
\end{abstract}

\begin{keyword}
Gamma-rays: bursts: accretion disks; black holes;  magnetic fields
\end{keyword}

\end{frontmatter}


\def\astrobj#1{#1}
\section{Introduction}
\label{sect:intro}

Gamma-ray bursts (GRBs) are among the most powerful explosion events in the universe since the
Big Bang (e.g. Kumar \& Zhang (2015) for a recent review). The very high luminosity is
generally believed to be produced in a relativistic jet with a large initial bulk Lorentz factor
($\Gamma_0$). However, it is unclear how such a jet is launched from the GRB central engine.

Two types of GRB central engine models have been discussed in the literature:
the hyper-accreting black hole (BH) models (e.g., Popham et al. 1999; Narayan et al. 2001;
Di Matteo et al. 2002; Kohri \& Mineshige 2002; Gu et al. 2006; Chen \& Beloborodov 2007;
Janiuk et al. 2007; Lei et al. 2009; Lei et al. 2013), and the millisecond magnetar models
(e.g. Usov 1992; Thompson 1994; Dai \& Lu 1998a,b; Wheeler et al. 2000; Zhang \& Meszaros 2001;
Metzger et al. 2008, 2011; Bucciantini et al. 2012). Recently, L{\"u} \& Zhang (2014) systematically
analyzed the Swift/XRT lightcurves of long GRBs, and found that both types of central engines may be
operating in GRBs. While some of them may habor a millisecond magnetar central engine, the majority
of GRBs likely have a hyper-accreting black hole central engine.

Within the black hole central engine scenario, there are two main energy reservoirs
to provide the jet power: the accretion energy in the disk that is mostly carried by neutrinos and
anti-neutrinos, which annihilate and power a bipolar outflow (Popham et al. 1999; Di Matteo et al. 2002);
and the spin energy of the BH,
which can be tapped by a magnetic field connecting the outer world through the
Blandford \& Znajek (1977, hereafter BZ) mechanism. It is unclear which one is
the main mechanism to launch the jet.

The luminosity and the initial Lorentz factor $\Gamma_0$ of the jet are two important parameters,
that carry the information about jet launching. Using the peak time of the early optical afterglow onset
bump as the deceleration time of the GRB ejecta, Liang et al. (2010) found a correlation between
$\Gamma_0$ and $E_{\gamma,\rm iso}$. The correlation is confirmed by later analyses
(e.g. Ghirlanda et al. 2012; L{\"u} et al. 2012; Liang et al. 2013). L\"u et al. (2012) also
discovered a somewhat tighter correlation between $\Gamma_0$ and the isotropic mean $\gamma$-ray
luminosity ($L_{\rm \gamma,iso}$).

Lei et al. (2013) investigated the baryon loading processses in the hyper-accreting black hole
central engine models for the two jet launching mechanisms: the non-magnetized $\nu \bar\nu$ -
annihilation mechanism and the strongly magnetized BZ mechanism. The two models have distinct
predictions between $\Gamma_0$ and the total jet power. Both of them predict a positive
correlation between $\Gamma_0$ and the power. The predicted slopes seem to be consistent with
the ones found by Liang et al. (2010) and L\"u et al. (2012). However, there are two missing links
between the model predictions and the observed correlations: First,
the central engine models directly predict the total jet power, whereas the previous studies only deal
with the observed $\gamma$-ray power. In order to get the total power, one needs to conduct an study
of the afterglow to retrieve the kinetic energy power. Second, the model predictions are relevant to
the beaming corrected total power, rather than the isotropic equivalent values presented in the prevous
studies. As a result, in order to confront the observational data with the central engine models, one
needs to collect a sample of GRBs whose $\Gamma_0$ can be measured, and whose $\gamma$-ray radiation
efficiency $\eta_\gamma$ and jet opening angle $\theta_{\rm j}$ can be constrained from the data.

In this paper, we work on a GRB sample whose $\Gamma_0$ values are measured using the peak of the early
optical afterglow light curve assuming that the peak time $T_{\rm p}$ corresponds to the blastwave deceleration time, and
whose jet opening-angle $\theta_{\rm j}$ are calculated from the jet break time $T_{\rm b}$ in afterglow light curves.
The kinetic energy $E_{\rm k,iso}$ of each GRB can be obtained by using the X-ray afterglow data during the normal
decay phase. Our goal is to compare the data with the black hole central engine model predictions. In order to test our predictions,
we also selected some bursts graded as ``Gold'' magnetar candidates by L{\"u} \& Zhang  2014 (see also Lyons et al. 2010
and Rowlinson et al. 2014) in our sample. We then
study the correlations between $\Gamma_0$ and the beaming corrected energies ($E_{\gamma}, E_{\rm K}, E_{\rm tot}$)
or mean luminosities ($L_{\gamma}, L_{\rm K}, L_{\rm tot}$), and compare the results with the predictions from the
$\nu \bar\nu$ - annihilation and the BZ jet launching mechanisms, respectively.
The data analysis results are presented in Section 2, and the theoretical implications are discussed in Section 3.
The results are summarized in Section 4.

\section{Sample Selection and Correlations}
\label{sect:tech}
We extensively search for the GRBs whose both initial Lorentz factor and jet opening angle can be constrained.
The initial Lorentz factor $\Gamma_0$ is obtained by the afterglow onset method, which uses the peak of the
early afterglow light curve to determine the deceleration time of blast wave. For a constant density medium case,
$\Gamma_0$ is given by (L{\"u} et al. 2012; Liang et al. 2013),
\begin{equation}
   \Gamma_{\rm 0}\simeq1.4\left[\frac{3E_{\rm \gamma,iso}(1+z)^{3}}{32{\pi}nm_{\rm p}c^{5}{\eta_{\gamma}}T_{\rm p}^{3}}\right]^{1/8},
   \label{eq:dec}
\end{equation}
where $n$ is the medium number density, $z$ is the redshift, $m_{\rm p}$ is the proton mass, $\eta_{\gamma}$ is the
efficiency of GRBs, $T_{\rm p}$ is peak time of the afterglow bump, and the numerical factor 1.4 is derived
from a more precise integration of blastwave dynamics before the deceleration time.

There are other methods for constraining $\Gamma_0$, such as the pair opacity constraint method (e.g. Lithwick \& Sari 2001)
and early external forward emission method (e.g. Zou \& Piran 2010), which can usually only give a lower (or upper) limit to
the Lorentz factor (L{\"u} et al. 2012). Therefore we generally focus on a sample of GRBs with $T_{\rm p}$ observed.
Liang et al. (2010) worked on an optical afterglow onset sample and constrained $\Gamma_0$ for about 20 GRBs. In this paper, we
worked on an expanded optical sample and also constrained $\Gamma_0$ with the same method. Most of the optical sample are taken
from Liang et al. (2010), and our entire sample includes 36 GRBs as presented in Table 1.

In order to identify jet breaks, one needs to systematically explore temporal breaks in the afterglow lightcurves. Theoretically
there are several types of temporal breaks (e.g. Zhang 2007), but observationally the observed breaks generally fall into two types.
The first type connects a shallow decay (decay index shallower than -1) segment to a normal decay (decay slope $\sim -1$) segment
(Zhang et al. 2006). This break is best interpreted as the cessation of energy injection into to the blastwave (Dai \& Lu 1998a,b;
Rees \& M\'esz\'aros 1998; Zhang \& M\'esz\'aros 2001; Liang et al. 2007), probably from a spinning-down millisecond magnetar
(Troja et al. 2007; Lyons et al. 2010; Rowlinson et al. 2013; L{\"u} \& Zhang 2014; Gompertz et al. 2014; L\"u et al. 2015). The second type is a transition from the normal decay phase (decay slope $\sim -1$) to a steeper phase (decay slope $\sim -2$),
which is best interpreted as a jet break (Rhoads 1999; Sari et al. 1999; Frail et al. 2001; Wu et al. 2004; Liang et al. 2008; Racusin et al. 2009). Differentiation of the two types of breaks is usually straightforward, but sometimes can be more complicated
(see Wang et al. 2015 for a recent detailed treatment).
In our work, we identify jet breaks from afterglow light curves that show a transition from the normal decay phase (decay slope $\sim -1$) to a steeper phase
(decay slope $\sim -2$). We identified 17 jet breaks in our $\Gamma_0$ sample, most of the values of $T_b$ are calculated from X-ray afterglow light curves, which are obtained from the UK Swift/XRT website \footnote{http://www.swift.ac.uk/xrt\_curves/} (Evans et al. 2007, 2009). The relevant information (frequency band, $\theta_j$, and references) is collected in Table 2. For these GRBs, we derive the jet opening angles using (e.g. Frail et al. 2001; Wu et al. 2004; Yi et al. 2015)
\begin{eqnarray}
\theta_{\rm j}=0.076 \left(\frac{T_{\rm b} }{1\ \rm day} \right)^{3/8}  \left(\frac{1+z}{2} \right)^{-3/8}\nonumber \\ \times
E_{\gamma,{\rm iso},53}^{-1/8}  \left(\frac{\eta_\gamma}{0.2} \right)^{1/8}
 \left(\frac{n }{1\ \rm cm^{-3}} \right)^{1/8}.
 \label{eq:theta_j}
\end{eqnarray}

For those GRBs without a jet break feature, we take the last observational point
of the normal phase as a lower limit of the jet break time, and derive a lower limit of the half
opening angle $\theta_{\rm j}$. We note that both Eq.(\ref{eq:dec}) and (\ref{eq:theta_j}) are derived
assuming a constant density medium. Long GRBs originate from the collapse of the massive stars, the
circumburst medium of long GRBs may be a wind type. However, according to Schulze et al. (2011), they
considered a group of GRBs with the lightcurve and SED analysis, the majority of cases consistent an
ISM type medium. Therefore, we simply assume an ISM case in this paper, having in mind that this is only a
rough approximation of real conditions.

With the half opening angle derived, we can perform beaming correction for energy and luminosity from isotropic values, i.e.
\begin{eqnarray}
 E & = & E_{\rm iso} f_{\rm b}  \\
 L & = & L_{\rm iso} f_{\rm b},
\end{eqnarray}
where
\begin{equation}
 f_{\rm b} = 1-\cos\theta_{\rm j} \simeq \theta_{\rm j}^2/2.
\end{equation}

Table 1 show the basic parameters of the GRBs in our sample, including name, redshift, $T_{90}$, and all the
isotropic values (e.g. $\gamma$-ray energy $E_{\rm \gamma,iso}$ and average $\gamma$-ray luminosity during the
prompt emission phase $L_{\rm \gamma,iso} = E_{\rm \gamma,iso}/T_{90}$). The derived parameters such as $\Gamma_0$
and $\theta_{\rm j}$ are presented in Table 2, along with beaming corrected values of energies and luminosities (e.g. $E_\gamma$ and $L_\gamma$).

We plot $\Gamma_0$ versus  $E_{\gamma}$ and $L_{\gamma}$ (blue and red data points) in the top and bottom panels
of Fig.1, respectively. The blue dots are the GRBs with $\theta_{\rm j}$ measurements, while red triangles have lower
limits of $\theta_{\rm j}$. For comparison, we also plot $\Gamma_0$ versus $E_{\rm \gamma,iso}$ and $L_{\rm \gamma,iso}$ (black dots).

We confirm the correlations of $\Gamma_0$-$E_{\rm \gamma,iso}$ and $\Gamma_0$-$L_{\rm \gamma,iso}$ . The best linear fitting results are
\begin{equation}
\log \Gamma_0=(-8.78\pm 2.53)+(0.21\pm 0.05) \log E_{\rm \gamma,iso},
\end{equation}
with a Pearson correlation coefficient $r=0.58$ and chance probability $p < 10^{-4}$, and
\begin{equation}
\log \Gamma_0=(-9.73\pm 2.08)+(0.23\pm 0.04) \log L_{\rm \gamma,iso},
\end{equation}
with $r=0.68$ ($p < 10^{-4}$).
These results are consistent with the results of Liang et al. (2010) and L\"u et al. (2012).

We next evaluate possible correlation between $\Gamma_0$ and beaming corrected energy/luminosity. Discarding the lower limits, the 17 GRBs with precise beaming corrections give rise to the correlations
\begin{equation}
\log \Gamma_0=(-3.78\pm 2.79)+(0.12\pm 0.06) \log E_{\rm \gamma},
\end{equation}
with $r=0.49$ ($p=0.05$), and
\begin{equation}
\log \Gamma_0=(-9.25\pm 3.16)+(0.24\pm 0.06) \log L_{\rm \gamma},
\end{equation}
with $r=0.69$ ($p = 2.39 \times 10^{-3}$).
These correlations are less tight compared with the correlations with the isotropic values.

It is generally accepted that GRB emission originates from two distinct locations. The prompt emission is
believed to be generated from an ``internal'' location before the ejecta is decelerated by the ambient
medium. The process dissipates the kinetic energy of the outflow through internal shocks (e.g. Rees
\& M{\'e}sz{\'a}ros 1994; Piran 1999) or dissipate magnetic energy in a Poynting flux dominated outflow
(e.g. Zhang \& Yan 2011). After the dissipation, the bulk of the energy is contained in the kinetic
form of the ejecta, which drives a relativistic shock into the medium to power the afterglow
(M{\'e}sz{\'a}ros \& Rees 1997; Sari et al. 1998; Wu et al. 2005; Yi et al. 2013, 2014, see
Gao et al. 2013 for a recent review).

Therefore the total energy of the relativistic outflow is a sum of two components,
the released $\gamma$-ray energy ($E_\gamma$) during the prompt emission phase, and the kinetic energy
($E_{\rm K}$) of blastwave during the afterglow phase. The GRB radiation efficiency is defined as (Lloyd-Ronning \& Zhang 2004),
\begin{equation}
\eta_\gamma = \frac{E_{\rm \gamma,iso}}{E_{\rm \gamma,iso}+E_{\rm K,iso}}
= \frac{E_\gamma}{E_\gamma+E_{\rm K}}= \frac{L_\gamma}{L_\gamma+L_{\rm K}}.
\end{equation}
The isotropic kinetic energy $E_{\rm K,iso}$ can be measured from the X-ray afterglow flux during the normal decay
phase. Since we assume that our
sample only includes GRB afterglows with constant density mediums, we calculate $E_{\rm K,iso}$
following Zhang et al. (2007). For $\nu>{\rm max}(\nu_{\rm m},\nu_{\rm c})$, we have
\begin{eqnarray}
E_{\rm K,iso,52
} & = & \left[\frac{\nu F_\nu (\nu=10^{18}~{\rm Hz})}{5.2\times
10^{-14} ~{\rm ergs~s^{-1} ~cm^{-2}} }\right]^{4/(p+2)} \nonumber \\ &\times
&D_{28}^{8/(p+2)}(1+z)^{-1}
t_{\rm d}^{(3p-2)/(p+2)}\nonumber \\
& \times & (1+Y)^{4/(p+2)} f_p^{-4/(p+2)}\epsilon_{\rm B,-2}^{(2-p)/(p+2)} \nonumber \\
& \times &\epsilon_{e,-1}^{4(1-p)/(p+2)} \nu_{18}{^{2(p-2)/(p+2)}};
\nonumber \\
\end{eqnarray}
and for the $\nu_m < \nu < \nu_c$, we have
\begin{eqnarray}
E_{\rm K,iso,52} & = & \left[\frac{\nu F_\nu (\nu=10^{18}~{\rm Hz})}{6.5\times
10^{-13} ~{\rm ergs~s^{-1} ~cm^{-2}} }\right]^{4/(p+3)} \nonumber  \\& \times &
D_{28}^{8/(p+3)}(1+z)^{-1}
 t_{\rm d}^{3(p-1)/(p+3)}\nonumber \\
& \times &f_p^{-4/(p+3)} \epsilon_{\rm B,-2}^{-(p+1)/(p+3)}
\epsilon_{\rm e,-1}^{4(1-p)/(p+3)} \nonumber \\ & \times & n^{-2/(p+3)}
\nu_{18}{^{2(p-3)/(p+3)}}.
\nonumber \\
\end{eqnarray}
Here $\nu f_\nu(\nu=10^{18}{\rm Hz})$ is the energy flux at $10^{18}$ Hz
(in units of ${\rm ergs~s^{-1} ~cm^{-2}}$), $D$ is the luminosity distance, $t_d$ is the time
in the observer frame in days, $Y$ is the Compton parameter, and $\varepsilon_{\rm e}$, $\varepsilon_{\rm B}$
are the equipartition parameters for electrons and magnetic fields in the shock, for which we adopt
the typical values 0.1, and 0.01, respectively. The parameter $f_p$ is a function of the power distribution
index $p$ (Zhang et al. 2007), i.e.
\begin{equation}
f_p \sim 6.73 \left(\frac{p-2}{p-1}\right)^{p-1} (3.3\times 10^{-6})^{(p-2.3)/2}.
\end{equation}
The above expressions are valid for $p>2$, which is relevant for all the GRBs in our sample.

The calculated the isotropic kinetic energy $E_{\rm K,iso}$, the total isotropic energy $E_{\rm tot,iso} =
E_{\rm \gamma,iso} + E_{\rm iso,K}$, the isotropic average kinetic power $L_{\rm K,iso} = E_{\rm K,iso}/T_{90}$,
the total isotropic average power $L_{\rm tot,iso}=E_{\rm tot,iso} / T_{90}$, as well as the radiation efficiency
$\eta_\gamma$ of all the GRBs in our sample are also presented in Table 1. The corresponding beaming corrected
energies ($E_{\rm K}$ and $E_{\rm tot}$) and powers ($L_{\rm K}$ and $L_{\rm tot}$) are also presented in Table 2.

We plot $\Gamma_0$ against the isotropic and beaming-corrected total energy and power in Fig.2. Best statistical fits give

\begin{equation}
\log \Gamma_0=(-8.24\pm 3.67)+(0.20\pm 0.07) \log E_{\rm tot,iso},
\end{equation}
with $r=0.42$ ($p=6.86 \times 10^{-3}$),
\begin{equation}
\log \Gamma_0=(-6.98\pm 2.73)+(0.18\pm 0.05) \log L_{\rm tot,iso},
\end{equation}
with $r=0.48$ ($p = 1.7\times 10^{-4}$),
\begin{equation}
\log \Gamma_0=(-2.08\pm 3.40)+(0.09\pm 0.07) \log E_{\rm tot},
\end{equation}
with $r=0.31$ ($p=0.22$), and
\begin{equation}
\log \Gamma_0=(-4.77\pm 4.06)+(0.14\pm 0.08) \log L_{\rm tot},
\label{Eq:Ltot}
\end{equation}
with $r=0.41$ ($p = 0.10$).

The correlations are not as tight as those of $\Gamma_0$-$E_{\gamma,\rm iso}$ ($E_{\gamma}$) and
$\Gamma_0$-$L_{\gamma,\rm iso}$ ($L_{\gamma}$), especially the $\Gamma_0 - E_{\rm tot}$ and
$\Gamma_0 - L_{\rm tot}$ relations. We note that there is a large uncertainty in estimating
$E_{\rm K,iso}$ and $E_{\rm K}$, due to the uncertainties in the shock parameters $\varepsilon_{\rm e}$
and $\varepsilon_{\rm B}$, so that there is a large uncertainty in $E_{\rm tot,iso}$ and $E_{\rm tot}$
as well. The fitting results are improved somewhat if one includes the GRBs with a lower limit of energy/luminosity
(red triangles in Figs.1 and 2). In any case, the relations between
$\Gamma_0$ and $E_{\rm tot}$ and $L_{\rm tot}$ allow us to diagnose the physical nature of the GRB central engine,
as will be discussed in the next section.

\begin{figure}
\includegraphics[angle=0,scale=0.30]{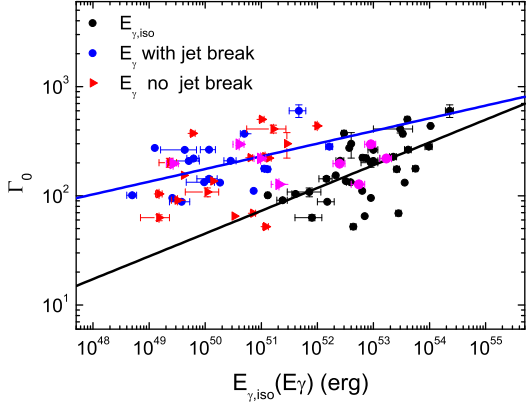}
\includegraphics[angle=0,scale=0.30]{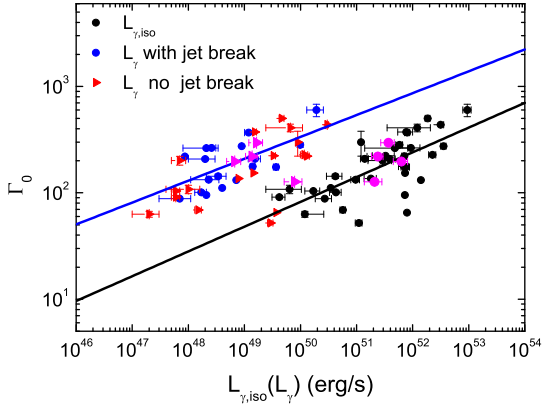}
\caption{Top panel: $\Gamma_0$ vs. $E_{\gamma}$; Bottom panel: $\Gamma_0$ vs. $L_{\gamma}$. The best-fitting results are $\Gamma_0 \propto E_{\gamma,iso}^{0.21}$ ($\propto E_{\gamma}^{0.12}$)and $\Gamma_0 \propto L_{\gamma,iso}^{0.23}$ ($\propto L_{\gamma}^{0.24}$). The blue dots are for the GRBs with beaming corrected energy/luminosity using the measured jet opening angle, while the red triangles indicate the GRBs with the lower limit of energy/luminosity. The isotropic energy/luminosity values are presented with black dots. The black and blue lines are the best fitting lines for the correlations. The magenta points and
triangles indicate the isotropic and beaming-corrected sample of the magnetar candidates.}
\end{figure}

\begin{figure}
\includegraphics[angle=0,scale=0.30]{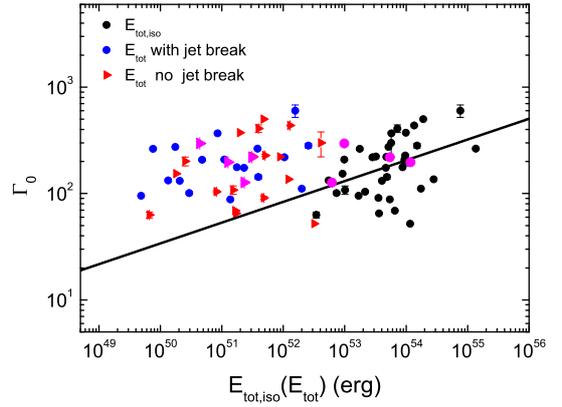}
\includegraphics[angle=0,scale=0.30]{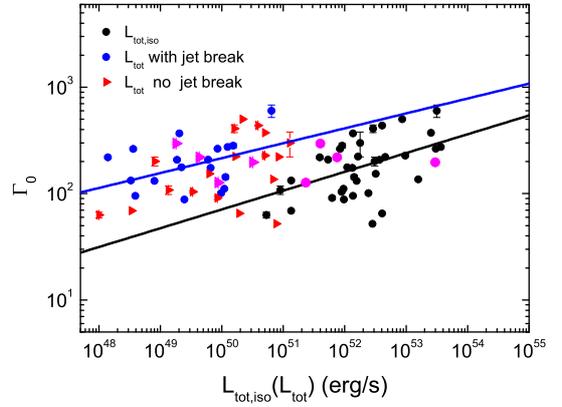}
\caption{$\Gamma_0$ vs. the total energy $E_{\rm tot}$ (top) and  the total luminosity $L_{\rm tot}$ (bottom). The best-fitting results are $\Gamma_0 \propto E_{\rm tot,iso}^{0.22}$ and  $\Gamma_0 \propto L_{\rm tot, iso}^{0.21}$ ($\propto L_{\rm tot}^{0.14}$). The symbols have the same meanings as in Figure 1.}
\end{figure}

\section{The BH Central Engine Model}
\label{sect:constraints}
The most popular model of GRB central engine invokes a stellar mass black hole surrounded by a hyper-accreting
disk (e.g., Popham et al. 1999; Narayan et al. 2001; Di Matteo et al. 2002; Kohri \& Mineshige 2002; Gu et al. 2006;
Chen \& Beloborodov 2007; Janiuk et al. 2007; Lei et al. 2009; Lei et al. 2013).
There are two main energy reservoirs to provide the jet power: the accretion energy in the
disk that is carried by neutrinos and anti-neutrinos, which annihilate and power a bipolar outflow; and the
spin energy of the black hole which can be tapped by a magnetic field connecting the outer world through the BZ mechanism.

Recently, Lei et al. (2013) investigated the baryon loading processes in the jet powered by these two
mechanisms, and calculated the initial Lorentz factor and total power of the jets for the two models.
With the data obtained in Sect. 2, we can now confront the model predictions with observations directly.

The neutrino annihilation ($\nu \bar{\nu} \rightarrow e^{+} e^{-}$) process can launch a relativistic jet
reaching the GRB
luminosity. For a system with black hole mass $M_\bullet$ and spin $a_\bullet$, the neutrino annihilation
power $\dot{E}_{\nu\bar{\nu}}$
depends on the accretion rate $\dot{M}$. The neutrino
annihilation power can be approximated as (Zalamea \& Beloborodov 2011),
\begin{equation}
\dot{E}_{\nu \bar{\nu}} \simeq 6.2 \times 10^{49} \left(\frac{R_{\rm ms} }{2} \right)^{-4.8} \left(\frac{m}{3}\right)^{-3/2} \dot{m}_{-1}^{9/4} ~{\rm erg \ s^{-1}},
\label{eq:Evv}
\end{equation}
where $m=M_{\bullet}/M_{\odot}$, $\dot{m} = \dot{M}/M_{\odot} \rm s^{-1}$, and $R_{\rm ms} = r_{\rm ms}/r_{\rm g}$
is the radius of the marginally stable orbit normalized to $r_{\rm g} =G M_{\odot} /c^2$.

Neutrino heating in the atmosphere just above the disk surface results in mass-loss from the hyperaccreting disk. The
dominant heating processes are electron neutrino absorption on baryons ($p+\bar{\nu}_e \rightarrow n+e^+$ and $n+\nu_e \rightarrow p+e^-$).

The baryon loading rate of the jet can be estimated as
\begin{eqnarray}
\dot{M}_{\rm j, \nu\bar{\nu}}
& = & 7.0 \times 10^{-7} A^{0.85} B^{-1.35} C^{0.22} \theta_{\rm j,-1}^2  \alpha_{-1}^{0.57} \epsilon_{-1}^{1.7} \nonumber \\
& & \left( \frac{R_{\rm ms} }{2} \right) ^{0.32} \dot{m}_{-1}^{1.7} \left(\frac{m}{3}\right)^{-0.9} \left(\frac{\xi}{2} \right)^{0.32} \ M_{\odot} {\rm s}^{-1}.
\end{eqnarray}
where $\xi \equiv r/r_{\rm ms}$ is the disk radius in terms of $r_{\rm ms}$, and $\alpha$ is the viscous parameter.
Here $\epsilon \simeq (1-E_{\rm ms})$ is the neutrino emission efficiency, and $E_{\rm ms}$ is the specific energy
corresponding to $r_{\rm ms}$.

The relativistic correction factors for a thin accretion disk around a Kerr BH are given by Riffert {\&} Herold (1995), i.e.
\begin{eqnarray}
\label{KerrFactors}
 A & = & 1 - 2R^{-1} + a_\bullet^2 R^{-2}, \\
 B & = & 1 - 3R^{-1} + 2a_\bullet R^{-3 / 2},  \\
 C & = & 1 - 4a_\bullet R^{-3 / 2} + 3 a_\bullet^2 R^{-2},
\end{eqnarray}
where $R=r/r_{\rm g}$ is the disk radius normalized to $r_{\rm g}$.

If most neutrino annihilation energy is converted into the kinetic energy of baryons after acceleration, the
jet would reach a Lorentz factor $\Gamma_{\rm max} = \eta$, with
\begin{eqnarray}
\eta  & \simeq   & \frac{\dot{E}_{\nu \bar{\nu}}}{\dot{M}_{\rm j, \nu\bar{\nu}} c^2} \nonumber \\
& =  & 50 A^{-0.85} B^{1.35} C^{-0.22} \theta_{j,-1}^{-2} \alpha_{-1}^{-0.57} \epsilon_{-1}^{-1.7} \left(\frac{\xi}{2}\right)^{-0.32} \nonumber \\
& \times & \left(\frac{ R_{ms}}{2}\right)^{-5.12} \left(\frac{m}{3}\right)^{-0.6} \dot{m}^{0.55}
\label{eq:eta}
\end{eqnarray}

\begin{figure}
\includegraphics[angle=0,scale=0.3]{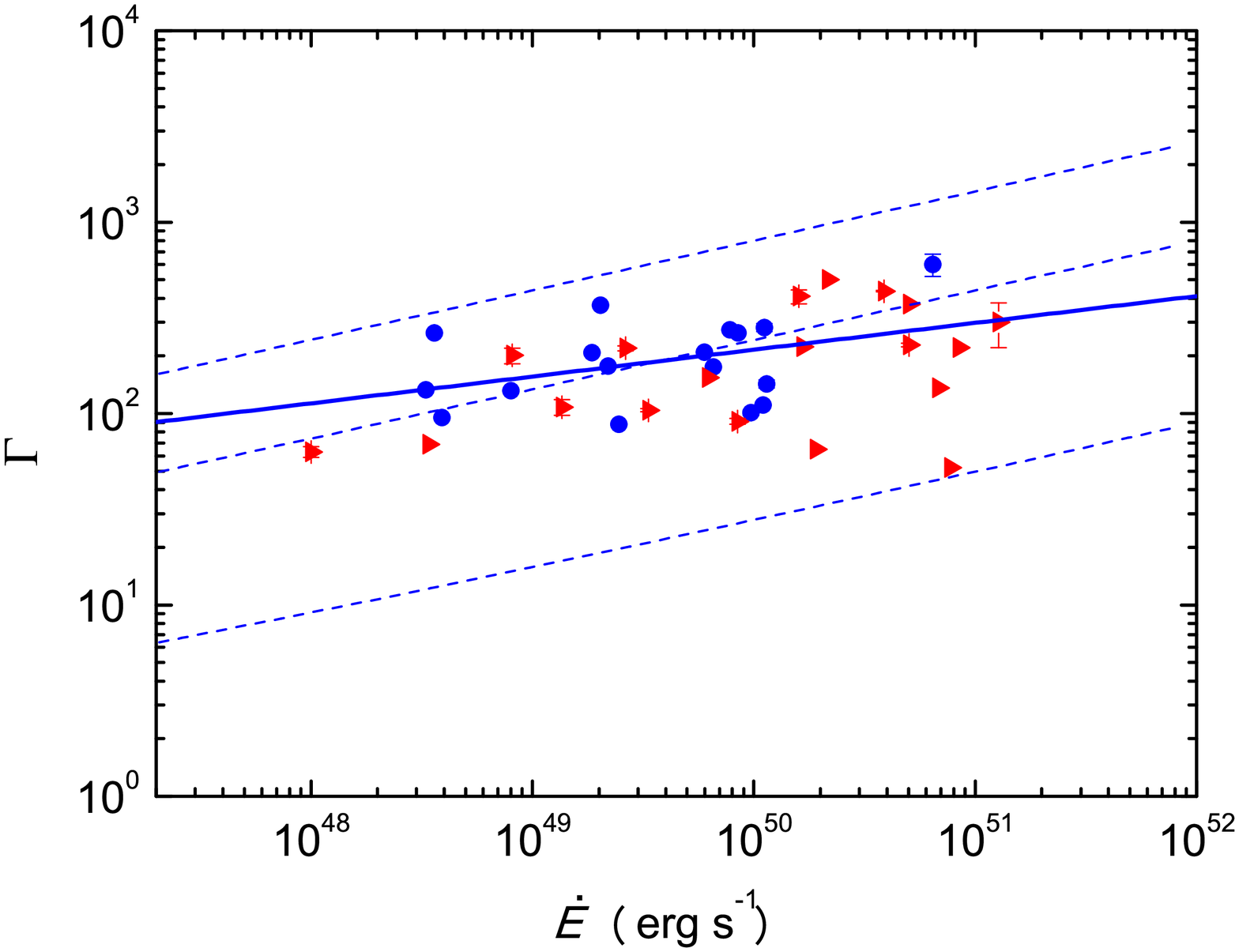}
\includegraphics[angle=0,scale=0.3]{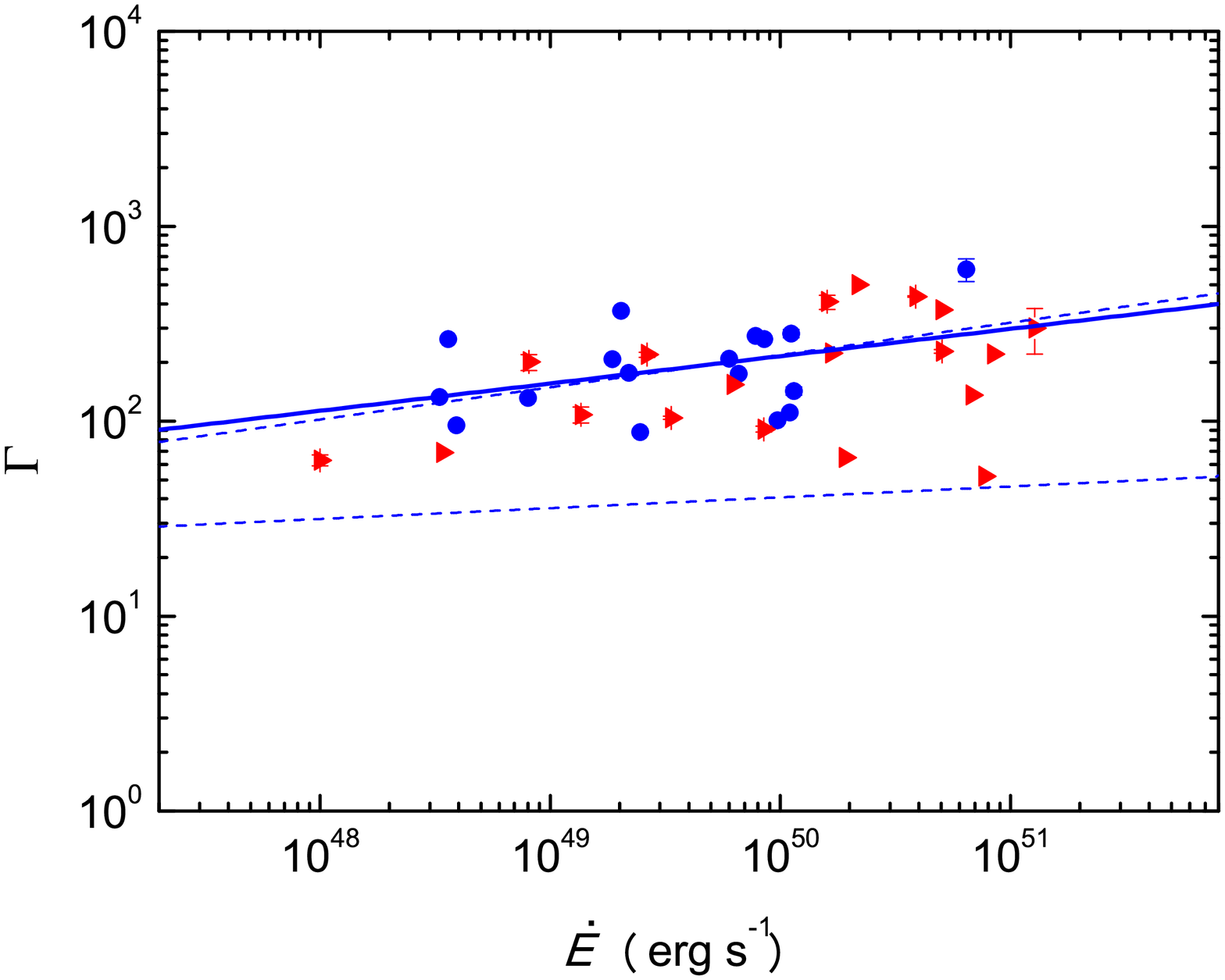}
\caption{A comparison between the observations and the predictions from the non-magnetized $\nu \bar\nu$ - annihilation mechanism (top) and the strongly magnetized BZ mechanism (bottom). The blue dots indicate GRBs with the beaming corrected luminosity, whereas red triangles are the ones with only lower limit. \textbf{Top panel}: The dashed lines show $\eta(\dot{m}) - \dot{E}_{\nu\bar{nu}} (\dot{m})$ relation for $\theta_{\rm j}$ = 0.022 (top), 0.04(middle) and 0.12 (bottom), respectively. Other parameters are fixed to $a_\bullet=0.998, m=3, \alpha=0.1$.
\textbf{Bottom panel}: The dashed line shows the $\Gamma_{\rm max} (\dot{m})$ - $\dot{E}_{\rm B} (\dot{m})$ relation, in which  we vary $\dot{m}$ in a wide range but fix the other parameters ($a_\bullet = 0.2, m=3, \theta_{\rm j}=0.2, \alpha=0.1$).  The $\Gamma_{\rm min} (\dot{m})$ - $\dot{E}_{\rm B} (\dot{m})$ is plotted with the dotted line, with $a_\bullet = 0.998, m=3, \theta_{\rm j}=0.01, \alpha=0.1$ adopted. In both panels, the best-fit to the data $\Gamma_0 \propto L_{\rm tot}^{0.14}$ are shown with a sold line for comparison. The magenta triangles indicate the magnetar candidates, including GRBs 060605, 060607A, 070110 and 081203A.}
\label{figCE}
\end{figure}

Inspecting Eqs. (\ref{eq:Evv}) and (\ref{eq:eta}), one finds that both $\dot{E}_{\nu \bar{\nu}}$ and $\eta$
are functions of $\dot{m}$, $a_\bullet$ and $m$. Usually the BH mass $m$ varies in the range of (3, 10).
Considering the hyper-accreting process during the prompt emission phase, the BH is quickly spun up, so that
the $a_\bullet$ dependence is not significant (Lei et al. 2013). As suggested by Lei et al. (2013),
the $\dot{m}$-dependence may be the key to define the $\eta - \dot{E}_{\nu \bar{\nu}}$ correlation, i.e.,
$\eta \propto \dot{E}_{\nu \bar{\nu}}^{0.26}$ (dashed line in Fig.\ref{figCE}a). Compared with the data,
the predicted index 0.26 (dashed lines) is larger than the observational value 0.14 (solid line) in Eq.(\ref{Eq:Ltot}).

As shown in Eq.(\ref{eq:eta}), the typical value of $\Gamma_{\rm max}$ is only $\sim 50$ for a typical jet
opening angle $\theta_{\rm j} = 0.1$, which is too small to be consistent with the GRB data. In Fig.\ref{figCE}a,
we plot $\Gamma_{\rm max} (\dot{m})$ versus $\dot{E}_{\nu \bar{\nu}} (\dot{m})$ with dashed lines for different
jet opening angle values, $\theta_{\rm j}$ = 0.022 (top), 0.04 (middle) and 0.12 (bottom). For each $\theta_{\rm j}$ case,
we vary $\dot{m}$ in a wide range of values, and keep other parameters to fixed values, e.g. $a_\bullet=0.998, m=3, \alpha=0.1$.
One can see that in order to account for the observations, one needs to invoke a relatively small opening angle.
The required opening angle is smaller than the typical measured values (Frail et al. 2001; Liang et al. 2008;
Racusin et al. 2009; Wang et al. 2015). The required small opening angle is also inconsistent with the predicted
value from numerical simulations of a $\nu - \bar\nu$-annihilation-powered jet
(e.g. Aloy, Janka \& Muller 2005; Harikae et al. 2010).

Another uncertainty for the $\nu\bar\nu$-annihilation model comes from the viscous parameter $\alpha$. In order
to reach a large $\Gamma$, one needs to invoke a relatively small $\alpha$. To see this, we calculate $\alpha$ and
$\dot{m}$ by equating $\Gamma_{\rm max} = \Gamma_0$ and $\dot{E}_{\nu \bar{\nu}} = L_{\rm tot}$, in which we take the
values of $\theta_{\rm j}$ from Table 1 and the maximum BH spin as $a_\bullet = 0.998$.
For most GRBs, e.g., 990123, 050820A, 061121, 070318, 071010B, 071031, 100901A, and 100906A, the required values
of $\alpha < 0.01$ are extremely small (King et al. 2007). This is another issue for the  $\nu\bar\nu$-annihilation model.

We now consider the prediction of the BZ model
(Blandford \& Znajek 1977). The BZ jet power from a BH with mass $M_{\bullet}$ and
angular momentum $J_\bullet$ is (Lee et al. 2000; Li 2000; Wang et al. 2002; McKinney 2005; Lei et al. 2005;
Lei et al. 2008; Lei \& Zhang 2011)
\begin{equation}
\dot{E}_{\rm B}=1.7 \times 10^{50} a_{\bullet}^2 m^2
B_{\bullet,15}^2 F(a_{\bullet}) \ {\rm erg \ s^{-1}},
\label{eq_Lmag}
\end{equation}
where $B_{\bullet,15}=B_{\bullet}/10^{15} {\rm G}$, $ F(a_{\bullet})=[(1+q^2)/q^2][(q+1/q) \arctan q-1]$, and $q= a_{\bullet} /(1+\sqrt{1-a^2_{\bullet}})$. Considering the balance between the magnetic pressure on the horizon and the ram
pressure of the innermost part of the accretion flow (e.g. Moderski et al. 1997),  one can then estimate the
magnetic field strength threading the BH horizon as
\begin{equation}
B_{\bullet} \simeq 7.4 \times 10^{16} \dot{m}^{1/2} m^{-1} (1+\sqrt{1-a_\bullet^2})^{-1} \rm{G}.
\end{equation}

Inserting it to Equation (\ref{eq_Lmag}), we obtain the magnetic power as a function of mass accretion rate and BH spin, i.e.
\begin{equation}
\dot{E}_{\rm B}=9.3 \times 10^{52} a_\bullet^2 \dot{m}_{-1}  X(a_\bullet) \ {\rm erg \ s^{-1}} ,
\end{equation}
where $X(a_\bullet)=F(a_\bullet)/(1+\sqrt{1-a_\bullet^2})^2 $.

The baryon-loading rate in a BZ jet is given in Lei et al. (2013) as
\begin{eqnarray}
\dot{M}_{\rm j, BZ}
& \simeq & 3.5 \times 10^{-7} A^{0.58} B^{-0.83} F_{\rm p,-1}^{-0.5} \theta_{\rm j,-1} \theta_{\rm B,-2}^{-1} \nonumber \\
& &  \alpha_{-1}^{0.38} \epsilon_{-1}^{0.83} \dot{m}_{-1}^{0.83} \left(\frac{m}{3}\right)^{-0.55}  r_{z,11}^{0.5} \ M_{\odot} {\rm s}^{-1},
\end{eqnarray}
where $F_{\rm p}$ is the fraction of protons, $r_{\rm z}$ is the distance from the BH in the jet direction, which is
normalized to $10^{11}$cm. Because of the existence of a strong magnetic field, protons with an ejected direction larger
than $\theta_{\rm B}$ with respect to the field lines would be blocked.

For a magnetized central engine, one can define a parameter
\begin{eqnarray}
\mu_{0} & \simeq &  \frac{\dot{E}_{\rm B} }{ \dot{M}_{\rm j,BZ} c^2} \nonumber \\
& = & 1.5 \times 10^5 A^{-0.58} B^{0.83} F_{\rm p,-1}^{0.5} \theta_{\rm j,-1}^{-1}  \theta_{\rm B,-2}  \alpha_{-1}^{-0.38} \epsilon_{-1}^{-0.83}    \nonumber \\
& & a_\bullet^2   X(a_\bullet)  \left(\frac{m}{3}\right)^{-0.55} \dot{m}_{-1}^{0.17} r_{z,11}^{-0.5},
\label{eq:mu}
\end{eqnarray}
This parameter denotes the maximum available energy per baryon in the jet.

The acceleration behavior of the jet is subject to uncertainties. Generally, the jet will reach a terminating Lorentz factor $\Gamma$ that satisfies
\begin{equation}
 \Gamma_{\rm min} < \Gamma < \Gamma_{\rm max},
\label{gamma}
\end{equation}
with the explicit value depending on the detailed dissipation process, such as the Internal-Collision induced Magnetic
Reconnection and Turbulence (ICMART) avalanche to discharge the magnetic energy (Zhang \& Yan 2011)\footnote{If the central
engine launches a two-component jet, another possible mechanism to dissipate magnetic energy is the shearing interaction
between the two jets (e.g. Wang et al. 2014). }. Following Lei et al. (2013), we take $\Gamma_{\rm min}=\max(\mu_{0}^{1/3},\eta)$
($\eta=\dot{E}_{\nu \bar{\nu}} /(\dot{M}_{\rm j, BZ} c^2)$) and $\Gamma_{\rm max} =  \mu_0$, which correspond to the beginning
and the end of the slow acceleration phase in a hybrid outflow, respectively (see Gao \& Zhang 2015 for a detailed discussion
of the acceleration dynamics of an arbitrarily magnetized relativistic or hybrid jet).

Based on Eqs. (\ref{eq_Lmag}) and (\ref{eq:mu}), the BZ power and mass loss rate are related to $\dot{m}$ as
$\dot{E}_{\rm BZ} \propto \dot{m}$ and $\mu_0 \propto \dot{m}^{0.17}$, respectively. Combining these dependents,
one derives $\mu_0 \propto \dot{E}_{\rm BZ}^{0.17}$ (dashed line in Fig.\ref{figCE}b), which agrees well with the
statistical correlation (\ref{Eq:Ltot}) (solid line in Fig.\ref{figCE}b).

In Fig.\ref{figCE}b, we plot $\Gamma_{\rm max}$ and $\Gamma_{\rm min}$ versus $\dot{E}_{\rm B} $. The dashed line
shows the $\Gamma_{\rm max} (\dot{m})$ - $\dot{E}_{\rm B} (\dot{m})$ relation, in which  we let $\dot{m}$ vary in a
wide range but fix the other parameters to typical values ($a_\bullet = 0.2, m=3, \theta_{\rm j}=0.2, \alpha=0.1$).
As discussed above, this dashed line gives the right slope to interpret the empirical $\Gamma_0 - L_{\rm tot}$ relation.
The values of $\Gamma_{\rm max}$ and $\dot{E}_{\rm BZ}$ can account for the observational $\Gamma_0$ and $L_{\rm tot}$
for most GRBs. In the same way, we plot $\Gamma_{\rm min} (\dot{m})$ versus $\dot{E}_{\rm B} (\dot{m})$ with the dotted line
in Fig.\ref{figCE}b, in which $a_\bullet = 0.998, m=3, \theta_{\rm j}=0.2, \alpha=0.1$ are adopted. The $\Gamma_{\rm min}$
value is significantly lower than the observations even for the fastest spinning BH. This suggests that the energy dissipation
region is far above the slow acceleration radius (Gao \& Zhang 2015), which is consistent with having significant magnetic
dissipation at a large emission radius, as is expected in the ICMART model (Zhang \& Yan 2011), and the requirement to interpret
GRB spectra within the fast-cooling synchrotron radiation model (Uhm \& Zhang 2014).

Another consideration is the dependence of BH spin, which may also introduce a correlation between $\Gamma$ and $\dot{E}_{\rm BZ}$.
However, with the consideration of the spin-up process due to accretion and the spin-down process due to the BZ process, the BH spin
parameter always evolves to an equilibrium value (Lei et al. 2005), so that the $a_\bullet$-dependence essentially does not enter
the problem.
Observationally our correlation was obtained from the average $\Gamma_0$ and $L$. In order to compare with the data one needs
to calculate the average $\Gamma$ and $\dot{E}_{\rm BZ}$ (Lei et al. 2013), which smears the $a_\bullet$ dependence.

\section{Discussion and Conclusions}
\label{sect:Conclusion}
In this paper, we worked on a GRB sample with measured initial Lorentz factor $\Gamma_0$ and the jet
opening-angle $\theta_{\rm j}$. We confirmed several previously discovered correlations (Liang et al.
2010; L\"u et al. 2012), and found some rough correlations between $\Gamma_0$ and the beaming
corrected $\gamma$-ray energy ($E_{\gamma}$) and the total ($\gamma$-ray plus kinetic) energy
($E_{\rm tot}$), and with the corresponding average luminosities ($L_{\gamma}, L_{\rm tot}$). Since the bulk Lorentz factor and the total power of the jet are directly related to the GRB
central engine, we can use the data to constrain the jet launching mechanism in the BH central engine
model.

Comparing the observational results with the theoretical predictions from the two types of BH central
engine models, i.e.,the $\nu \bar\nu$ - annihilation mechanism (non-magnetized model) and the BZ mechanism
(strongly magnetized model), we found that the required parameters are contrived for the former mechanism,
whereas the latter (BZ) mechanism can generally account for the observations. We tentatively suggest the GRB
jet composition contains a significant fraction of Poynting flux, and the GRB prompt emission is likely powered
by dissipation of magnetic field energy (which also seen Zhang \& Yan 2011; Yi et al. 2016).

Two caveats need to be mentioned. First, the $\Gamma_0 - E_{\rm tot}$ and $\Gamma_0 - L_{\rm tot}$ correlations
are very rough, with a lower correlation coefficient than the  $\Gamma_0 - E_{\gamma}$ and $\Gamma_0 - L_{\gamma}$
relations. There are large uncertainties involved when constraining $E_{\rm K}$ and $L_{\rm K}$, due to the uncertainties
in the values of the shock equipartition parameters. On the other hand, the wide scatter is not in conflict with the
theoretical models, which predicts a wide range of $\Gamma_0$ and $L_{\rm tot}$ due to their dependences on many model
parameters (e.g. $\dot m$, $m$, $a_\bullet$, $\alpha$, $\theta_{\rm j}$). In any case, the available data allow us to
reach the conclusion that the data are more consistent with the BZ mechanism than the non-magnetized $\nu\bar\nu$-annihilation
mechanism.

Second, we focus on the BH central engine model. Another frequently discussed central engine model is the millisecond magnetar
central engine model (e.g. Metzger et al. 2011). We justify this approach by excluding the Gold-sample magnetar candidates as
identified by (L{\"u} \& Zhang 2014). However, there is no smoking-gun criterion for confirm that all the GRBs in our sample
have a BH central engine.

\section*{Acknowledgments}
We thank an anonymous referee for helpful suggestions and Hou-Jun L{\"u} for a comment. We also acknowledge the use of the public data from the UK Swift Science Data Center. This work is supported by National Basic Research Program (``973'' Program) of China under grant No. 2014CB845800, the program A for Outstanding PhD  candidate of Nanjing University, and under grant No. xkj201614 of Qufu Normal University, National Natural Science Foundation of China under grants U1431124, 11322328, 11361140349 (China-Israel jointed program), the One-Hundred-Talent Program (XFW) and the Strategic Priority Research Program The Emergence of Cosmological Structure (grant No. XDB09000000) of the Chinese Academy of Sciences.

%

\begin{table*}
\scriptsize{\tiny}
\centering
\caption{The Parameters of the GRBs in this Sample}
\begin{tabular}{cccccccccccc}
\hline		
\hline						
GRB &
$z ^{ref}$&
$T_{p,z}^{a}$ &
$T_{90}^{a}$ &
$E_{\gamma,\rm iso}^{b}$ &
$E_{\rm K,iso}^{c}$ &
$E_{\rm tot,iso}^{c}$ &
$L_{\gamma,\rm iso}^{d}$ &
$L_{\rm K,iso}^{d}$ &
$L_{\rm tot,iso}^{d}$ &
$\eta_\gamma$ ($\%$)&
Refs. \\
\hline	
With Jet Break\\
\hline
				
990123	&	1.61$^{(1)}$	&	19.16	$\pm$	5.75	&	63.3	$\pm$	0.3	&	22900	 $\pm$	3700	&	53400	&	76300	&	944	$\pm$	157	&	 2201.8	&	3146.02	&	30	 &	36, 37  \\
050820A	&	2.615$^{(2)}$	&	108.17	$\pm$	4.62	&	600		&	9700	$\pm$	 1400	&	5371.45	&	15071.45	&	58	$\pm$	8	&	32.36	 &	 90.81	&	64.4	 &	38, 39 \\
050922C	&	2.198$^{(3)}$	&	42			&	4.5		&	370		&	4772.5	&	5142.5	 &	356	$\pm$	39	&	3391.66	&	3654.6	&	7.2	&	38, 39 \\
060210	&	3.91$^{(4)}$	&	97			&	220	$\pm$	70	&	4150	$\pm$	570	&	 131322.61	&	135472.61	&	93	$\pm$	42	&	2930.88	&	 3023.5	&	3.1	&	38, 39 \\
060418	&	1.49$^{(5)}$	&	60.73	$\pm$	0.82	&	52	$\pm$	1	&	1000	 $\pm$	200	&	753.07	&	1753.07	&	48	$\pm$	34	&	36.06	 &	 83.95	&	57	&	 38, 39 \\
061121	&	1.314$^{(6)}$	&	250			&	81	$\pm$	5	&	2610	$\pm$	300	&	 2052.15	&	4662.15	&	75	$\pm$	13	&	58.63	&	133.19	 &	 56	&	40, 39 \\
070318	&	0.84$^{(7)}$	&	162.09	$\pm$	15.26	&	63	$\pm$	3	&	145	$\pm$	 38	&	4727.19	&	4872.19	&	4.2	$\pm$	1.3	&	138.06	&	 142.3	 &	3	&	38, 39 \\
070411	&	2.95$^{(8)}$	&	113.83	$\pm$	1.27	&	101	$\pm$	5	&	1000	 $\pm$	200	&	8365.96	&	9365.96	&	39	$\pm$	10	&	327.18	 &	 366.29	&	10.7	 &	38, 39 \\
071010A	&	0.98$^{(9)}$	&	185.95	$\pm$	12.31	&	6	$\pm$	1	&	13	$\pm$	 1	&	721.64	&	734.64	&	4.3	$\pm$	1	&	238.14	&	 242.43	&	1.8	&	38, 39 \\
071010B	&	0.947$^{(10)}$	&	67			&	35.7	$\pm$	0.5	&	255	$\pm$	41	&	 727.13	&	982.13	&	14	$\pm$	2.3	&	39.61	&	53.5	 &	 26	&	38, 39 \\
071031	&	2.692$^{(11)}$	&	275.88	$\pm$	0.42	&	150.5		&	390	$\pm$	60	 &	155.4	&	545.4	&	9.6	$\pm$	1.7	&	3.81	&	 13.38	 &	 71.5	&	38, 39 \\
080603A	&	1.688$^{(12)}$	&	595.24			&	150		&	150	$\pm$	50	&	5251.29	 &	5401.29	&	2.7	$\pm$	0.87	&	94.1	&	96.79	&	 2.8	 &	 36, 39  \\
100901A	&	1.408$^{(13)}$	&	523.26		&	439		&	630		&	16732.33	&	 17362.33	&	3.46		&	91.78	&	95.24	&	3.6	&	41, 42  \\
100906A	&	1.727$^{(13)}$	&	37	$\pm$	1.6	&	114.4	$\pm$	1.6	&	3340	$\pm$	 300	&	2381.73	&	5721.73	&	80	$\pm$	8.3	&	56.77	&	 136.39	&	58.4	&	40, 39  \\
110205A	&	2.22$^{(14)}$	&	311	$\pm$	25	&	257	$\pm$	25	&	5600	$\pm$	600	 &	3121.72	&	8721.72	&	70	$\pm$	14	&	39.11	&	 109.28	&	 64.2	&	40, 39 \\
120119A	&	1.728$^{(15)}$	&	364.0$\pm$	0.4	&	70	$\pm$	4		&	3600	&	417	 &	4017	&	140.3		&	16.3	&	156.6	&	90	&	 43 \\
120404A	&	2.876$^{(16)}$	&	642.7	$\pm$	1.8	&	48	$\pm$	16	&	900		&	780	 &	1680	&	72.7&	63.0	&	 135.7	&	54	&	44, 45  \\

\hline
No Jet Break\\
\hline
050730	&	3.97$^{(17)}$	&	120.09	$\pm$	27.76	&	155	$\pm$	20	&	900	$\pm$	 300	&	8612.23	&	9512.23	&	29	$\pm$	13	&	276.15	&	 305.01	&	9.5	& 38, 39 \\
060904B 	&	0.703$^{(18)}$	&	271.91	$\pm$	33.75	&	192	$\pm$	5	&	72	 $\pm$	43	&	940.31	&	1012.31	&	0.64	$\pm$	0.4	&	 8.34	 &	 8.98	&	 7.1& 38, 39 	\\
061007	&	1.262$^{(19)}$	&	34.62	$\pm$	0.18	&	75	$\pm$	5	&	10465	 $\pm$	694	&	2994.25	&	13459.25	&	316	$\pm$	42	&	 90.31	 &	 405.93	&	 77.8	& 38, 39 \\
070419A 	&	0.97$^{(20)}$	&	297.98	$\pm$	10.62	&	112	$\pm$	2	&	24	 $\pm$	5	&	3507.41	&	3531.41	&	0.42	$\pm$	0.11	&	 61.69	 &	62.11	&	 0.7	& 38, 39 \\
080129	&	4.394$^{(21)}$	&	2224.69			&	48	$\pm$	2	&	700		&	2913.81	 &	3613.81	&	79	$\pm$	0	&	327.44	&	406.1	&	 19.4	 & 36 \\
080319C 	&	1.95$^{(22)}$	&	117.38	$\pm$	3.22	&	29.55		&	2255	 $\pm$	335	&	7440.78	&	9695.78	&	225	$\pm$	33	&	742.82	 &	 967.94	&	23.3	 & 38, 39  \\
080330	&	1.51$^{(23)}$	&	247.77	$\pm$	6.79	&	61	$\pm$	9	&	41	$\pm$	 6	&	2109.23	&	2150.23	&	1.7	$\pm$	0.5	&	86.79	&	 88.48	 &	1.9	& 38, 39 \\
080710	&	0.845$^{(24)}$	&	1192.91	$\pm$	2.24	&	120	$\pm$	17	&	80	$\pm$	 40	&	264.51	&	344.51	&	1.2	$\pm$	1.4	&	4.07	&	 5.3	 &	 23.2	& 38, 39 \\
080810	&	3.35$^{(25)}$	&	27.02	$\pm$	0.26	&	108	$\pm$	5	&	3000	 $\pm$	2000	&	4185.19	&	7185.19	&	121	$\pm$	86	&	 168.57	&	 289.4	&	 41.8	& 38, 39 \\
081203A 	&	2.1$^{(26)}$	&	118.09	$\pm$	0.46	&	223		&	1700	$\pm$	 400	&	1122.61	&	2822.61	&	24	$\pm$	8.7	&	15.61	&	 39.24	 &	60.2	& 38, 39 \\
090313	&	3.375$^{(27)}$	&	242.51	$\pm$	35.2	&	78	$\pm$	19	&	320		&	 27685.23	&	28005.23	&	18	$\pm$	4.4	&	1552.86	&	 1570.81	&	1.1	&36  \\
090424	&	0.544$^{(28)}$	&	--			&	49.47	$\pm$	0.9	&	400		&	5312.15	 &	5712.15	&	12	$\pm$	0	&	165.8	&	178.28	&	7	 &39 \\
090812	&	2.452$^{(29)}$	&	17.38			&	75.9		&	4030	$\pm$	400	&	 14882.7	&	18912.7	&	185	$\pm$	18	&	676.88	&	860.17	 &	 21.3	&40, 39 \\
091024	&	1.092$^{(30)}$	&	1912			&	1020		&	2800	$\pm$	300	&	 3725.29	&	6525.29	&	5.7	$\pm$	0.62	&	7.64	&	 13.38	&	 42.9	&40, 39 \\
091029	&	2.752$^{(31)}$	&	88			&	39.2		&	740	$\pm$	74	&	4030.3	 &	4770.3	&	71	$\pm$	7.1	&	385.76	&	456.59	&	 15.5	 &40, 39  \\
100621A 	&	0.542$^{(32)}$	&	3443			&	63.6	$\pm$	1.7	&	437	$\pm$	 50	&	11175.96	&	11612.96	&	11	$\pm$	1.5	&	 270.96	&	 281.56	&	3.8	 &40, 39 \\
100728B 	&	2.106$^{(33)}$	&	16			&	12.1	$\pm$	2.4	&	300	$\pm$	30	 &	9566.5	&	9866.5	&	77	$\pm$	23	&	2455.67	&	 2532.67	 &	 3	&40, 39 \\
110213A 	&	1.46$^{(34)}$	&	81			&	48	$\pm$	6	&	640	$\pm$	60	&	 2575.27	&	3215.27	&	33	$\pm$	14	&	131.98	&	164.78	 &	 19.9	&40, 39 \\
120815A 	&	2.358$^{(35)}$	&	147.6	$\pm$	1.4&	9.7	$\pm$	2.5	&	212	&	708	 &	920	&	73.4	&	245.1	&	318.5	&	 23	& 45  \\
\hline	
\textbf{Magnetar Candidate$^{e}$}\\
\hline
060605	&	3.8	&	83.14	$\pm$	2.7	&	19	$\pm$	1	&	250	$\pm$	60	&	11500	 &	11750	&	63	$\pm$	18	&	2905.26	&	2968.42	&	0.02	 &	 38, 46	\\	
060607A	&	3.082	&	42.89	$\pm$	0.62	&	100	$\pm$	5	&	900	$\pm$	200	&	 82.2	&	982.2	&	37	$\pm$	10	&	3.36	&	40.09	&	 0.92	&	38, 46	\\	
070110	&	2.352	&	350			&	89	$\pm$	7	&	550	$\pm$	150	&	68.7	&	 618.7	&	21	$\pm$	7.3	&	2.59	&	23.30	&	0.89	&	 40, 46	\\	
081203A	&	2.1	&	118.09	$\pm$	0.46	&	223			&	1700	$\pm$	400	&	 3700	&	5400	&	24	$\pm$	5.6	&	51.43	&	75.07	&	 0.31	 &	38, 46	\\			
\hline					
\end{tabular}
\begin{flushleft}
$^{a}$ In units of seconds.\\
$^{b}$ The istropic energy (1-10$^{4}$ KeV in the burst frame, in units of $10^{50} erg$).\\
$^{c}$ In units of $10^{50}$ erg.\\
$^{d}$ In units of $10^{50}$ erg/s.\\
\textbf{$^{e}$ The GRBs of magnetar candidates are without the jet break detected.}\\
Refs: 1: Galama et al. (1999); 2: Ledoux et al.(2005); 3: Jakobsson et al.(2005); 4: Cucchiara et al. (2006); 5: Molinari et al. (2007); 6: Page et al.(2007); 7: Jaunsen et al. (2007); 8: Jakobsson et al.(2007); 9: Prochaska et al.(2007); 10: Cenko et al. (2007);
11: Ledoux et al. (2007); 12: Guidorzi et al.(2011); 13: Gorbovskoy et al.(2012); 14: Vreeswijk et al. (2011); 15: Cucchiara
\& Prochaska(2012); 16: Cucchiara(2012);
17: Rol et al. (2005); 18: Fugazza et al.(2006); 19: Mundell et al.(2007); 20: Melandri et al. (2009); 21:
Greiner et al.(2009); 22: Wiersema et al.(2008); 23: Cucchiara et al. (2008); 24: Perley et al. (2008); 25:
Prochaska et al.(2008); 26: Landsman et al.(2008); 27: de Ugarte Postigo et al.(2009a); 28: Chornock et al. (2009);
29: de Ugarte Postigo et al. (2009b); 30: Cucchiara et al. (2009); 31: Chornock et al.(2009); 32: Milvang-Jensen et al.(2010);
33: Flores et al. (2010); 34: Milne et al. (2011); 35: Malesani et al.(2012);
36: Melandri et al. (2010); 37: Amati et al. (2008); 38: Liang et al. (2010); 39: L\"u et al. (2012); 40: Ghirlanda et al. (2012); 41: Liang et al. (2013);
42: Gorbovskoy et al. (2012); 43: Morgan et al. (2014); 44: Guidorzi et al. (2014); 45: Kr{\"u}hler et al. (2013); 46: L\"u \& Zhang (2014). \\
\end{flushleft}
\end{table*}

\begin{table*}
\scriptsize
\centering
\caption{The Beaming Correction Parameters of the GRBs in this Sample}
\begin{tabular}{cccccccccccc}
\hline		
\hline						
GRB &
$\Gamma_{0}$ &
$T_{\rm b}^{a}$ &
$Band^{b}$ &
$\theta_{\rm j}^{c}$ &
$E_{\gamma}^{d}$  &
$E_{\rm K}^{d}$  &
$E_{\rm tot}^{d}$  &
$L_{\gamma}^{e}$  &
$L_{\rm K}^{e}$  &
$L_{\rm tot}^{e}$  &
Refs. \\
\hline	
With Jet Break\\
\hline
				
990123	&	600	$\pm$	80	&	2.04	$\pm$	0.46	&	O	&	0.064	$\pm$	0.005	 &	46.79	$\pm$	15.47	&	109.12	&	155.91	&	1.929	 $\pm$	 0.635	&	4.500	 &	6.428	&	1	\\
050820A 	&	282	$\pm$	14	&	27.53&	O/X	&	0.184	&	163.96	$\pm$	23.66	&	 90.79	&	254.76	&	0.988	 $\pm$	 0.015	&	0.547	&	 1.535	&	2	\\
050922C 	&	274		&	0.09	&	O/X	&	0.026	&	0.13	&	1.62	&	1.75	 &	0.089	 	&	1.152	&	1.242	&	2	\\
060210	&	264	$\pm$	4	&	0.33	$\pm$	0.07	&	X	&	0.024	$\pm$	0.002	 &	1.17	$\pm$	0.35	&	36.91	&	38.08	&	0.026	 $\pm$	 0.005	&	0.820	 &	0.850	&	3	\\
060418	&	263	$\pm$	7	&	0.07	$\pm$	0.04	&	X	&	0.029	$\pm$	0.006	 &	0.43	$\pm$	0.27	&	0.33	&	0.76	&	0.021	 $\pm$	 0.013	&	0.020	 &	0.036	&	4	\\
061121	&	175	$\pm$	2	&	2.31		&	X	&	0.099		&	12.89	$\pm$	 1.48	&	10.13	&	23.02	&	0.368	$\pm$	0.040	&	 0.290	 &	 0.658	&	5	 \\
070318	&	143	$\pm$	7	&	3.57	$\pm$	0.63	&	X	&	0.127	$\pm$	0.008	 &	1.17	$\pm$	0.46	&	38.00	&	39.17	&	0.034	 $\pm$	 0.013	&	1.110	 &	1.144	&	4	\\
070411	&	208	$\pm$	5	&	0.24	$\pm$	0.11	&	X	&	0.032	$\pm$	0.005	 &	0.51	$\pm$	0.28	&	4.25	&	4.76	&	0.020	 $\pm$	 0.010	&	0.170	 &	0.186	&	4	\\
071010A 	&	101	$\pm$	3	&	0.81	$\pm$	0.20	&	X	&	0.090	$\pm$	 0.008	&	0.05	$\pm$	0.01	&	2.90	&	2.95	&	 0.017	 $\pm$	0.004	&	 0.960	&	0.973	&	4	\\
071010B 	&	209	$\pm$	4	&	3.44	$\pm$	0.39	&	O	&	0.150	$\pm$	 0.006	&	2.85	$\pm$	0.70	&	8.13	&	10.98	&	 0.155	 $\pm$	0.038	&	 0.440	&	0.598	&	6	\\
071031	&	133	$\pm$	3	&	0.71	$\pm$	0.35	&	X	&	0.070	$\pm$	0.013	 &	0.96	$\pm$	0.50	&	0.38	&	1.34	&	0.023	 $\pm$	 0.012	&	0.010	 &	0.033	&	4	\\
080603A 	&	88		&	1.16$\pm$	0.46	&	O	&	0.071	$\pm$	0.011	&	 0.38	$\pm$	0.15	&	13.32	&	13.71	&	0.007	$\pm$	 0.004	 &	0.239	&	 0.246	&	7	\\
100901A	&	111		&	11.57		&	X	&	0.152		&	7.28		&	193.26	&	 200.54	&	0.040		&	1.060	&	1.100	&	8	\\
100906A 	&	369		&	0.59	$\pm$	0.05	&	X	&	0.055	$\pm$	0.002	&	 4.98	$\pm$	0.76	&	3.55	&	8.53	&	0.119	 $\pm$	 0.018	&	0.080	&	 0.203	&	8	\\
110205A 	&	177		&	1.20		&	O/X	&	0.064		&	11.29	$\pm$	1.21	 &	6.29	&	17.59	&	0.141	$\pm$	0.014	&	0.080	 &	 0.220	&	9	\\
120119A	&	132		&	0.13	$\pm$	0.02&	X	&	0.032$\pm$	0.002&	1.84&	0.21	 &	2.06	&	0.072		&	0.008	&	0.080	&	10	\\
120404A 	&	95		&	0.06	&	O	&	0.024		&	0.26	&	0.22	&	 0.48	&	0.021	&	0.018	&	0.039	&	11	\\

\hline
No Jet Break\\
\hline
050730	&	201	$\pm$	19	&	$>$0.12	&	X	&	$>$0.023	&	$>$0.23	$\pm$	0.08	 &	$>$2.19	&	$>$2.52	&	$>$0.007	$\pm$	0.002	&	 $>$0.070	 &	 $>$0.081	  	 \\
060904B 	&	108	$\pm$	10	&	$>$4.53	&	X	&	$>$0.174	&	$>$1.09	$\pm$	 0.65	&	$>$14.17	&	$>$15.32	&	$>$0.010	$\pm$	0.006	 &	 $>$0.126	&	 $>$0.136	 	\\
061007	&	436	$\pm$	3	&	$>$7.67	&	X	&	$>$0.138	&	$>$99.19	$\pm$	 6.58	&	$>$28.38	&	$>$128.16	&	$>$2.992	$\pm$	0.185	 &	 $>$0.856	 &	 $>$ 3.865	  	\\
070419A 	&	91	$\pm$	3	&	$>$6.88	&	X	&	$>$0.165	&	$>$0.32	$\pm$	 0.07	&	$>$47.50	&	$>$48.07	&	$>$0.006	$\pm$	0.001	 &	 $>$0.835	&	 $>$ 0.846	 	\\
080129	&	65		&	$>$4.65	&	X	&	$>$0.097	&	$>$3.30		&	$>$13.74	&	 $>$17.00	&	$>$0.371		&	$>$1.544	&	$>$1.910	 	 \\
080319C 	&	228	$\pm$	5	&	$>$4.02	&	X	&	$>$0.102	&	$>$11.70	$\pm$	 1.74	&	$>$38.61	&	$>$50.44	&	$>$1.168	$\pm$	 0.174	 &	 $>$3.854	&	 $>$5.035	 	\\
080330	&	104	$\pm$	2	&	$>$1.35	&	X	&	$>$0.087	&	$>$0.15	$\pm$	0.02	 &	$>$7.94	&	$>$8.14	&	$>$0.006	$\pm$	0.001	&	 $>$0.327	 &	 $>$0.335	  	 \\
080710	&	63	$\pm$	4	&	$>$0.22	&	X	&	$>$0.062	&	$>$0.15	$\pm$	0.08	 &	$>$0.51	&	$>$0.66	&	$>$0.002	$\pm$	0.001	&	 $>$0.008	 &	 $>$0.010	  	 \\
080810	&	409	$\pm$	34	&	$>$5.76	&	X	&	$>$0.105	&	$>$16.44	$\pm$	 10.96	&	$>$22.93	&	$>$39.61	&	$>$0.662	$\pm$	0.421	 &	 $>$0.924	 &	 $>$1.595	  	\\
081203A 	&	219	$\pm$	6	&	$>$3.95	&	X	&	$>$0.116	&	$>$11.42	$\pm$	 2.69	&	$>$7.54	&	$>$18.99	&	$>$0.159	$\pm$	0.037	 &	 $>$0.105	&	$>$ 0.264	 	\\
090313	&	136	$\pm$	0	&	$>$6.66	&	X	&	$>$0.093	&	$>$1.38		&	 $>$119.81	&	$>$121.11	&	$>$0.078		&	$>$6.720	&	 $>$6.793	 	 \\
090424$^{(f)}$	&	300	$\pm$	79	&	$>$58.33	&	X	&	$>$0.378	&	$>$28.64		 &	$>$380.39	&	$>$408.09	& $>$	0.894		&	 $>$11.872	 &	 $>$12.737	  	 \\
090812	&	501	&	$>$2.25	&	X	&	$>$0.071	&	$>$10.18	$\pm$	1.01	&	 $>$37.60	&	$>$47.67	&	$>$0.463	$\pm$	0.046	&	 $>$1.710	 &	 $>$2.168	  	 \\
091024	&	69		&	$>$0.97	&	X	&	$>$0.071	&	$>$7.16	$\pm$	0.77	&	 $>$9.52	&	$>$16.45	&	$>$0.015	$\pm$	0.002	&	 $>$0.020	 &	 $>$0.034	&	 \\
091029	&	221		&	$>$21.82	&	X	&	$>$0.192	&	$>$13.63	$\pm$	1.36	 &	$>$74.21	&	$>$87.93	&	$>$1.304	$\pm$	0.130	&	 $>$7.103	 &	 $>$8.416	  	 \\
100621A 	&	52		&	$>$20.55	&	X	&	$>$0.234	&	$>$12.00	$\pm$	 1.37	&	$>$306.77	&	$>$317.94	&	$>$0.291	$\pm$	0.032	 &	 $>$7.438	 &	 $>$7.709	 	 \\
100728B 	&	373		&	$>$1.18	&	X	&	$>$0.063	&	$>$0.60	$\pm$	0.06	&	 $>$18.98	&	$>$19.58	&	$>$0.153	$\pm$	0.012	&	 $>$4.872	 &	 $>$5.026	  	 \\
110213A 	&	223		&	$>$5.60	&	X	&	$>$0.142	&	$>$6.43	$\pm$	0.60	&	 $>$25.88	&	$>$32.42	&	$>$0.330	$\pm$	0.027	&	 $>$1.326	 &	 $>$1.661	  	 \\
120815A 	&	154		&	$>$0.57	&	X	&	$>$0.063	&	$>$0.41	&	$>$1.38	&	 $>$1.80	&	$>$0.143	&	$>$0.479	&	$>$0.622	 	 \\
\hline	
\textbf{Magnetar Candidate}\\
\hline
060605	&	197	$\pm$	6	&	$>$0.85	&	X	&	$>$0.046	&	$>$0.27	$\pm$	0.06	 &	$>$12.29	&	$>$12.56	&	$>$0.067	$\pm$	0.019	&	 $>$3.105	&	 $>$3.172	\\	
060607A	&	296	$\pm$	8	&	$>$2.12	&	X	&	$>$0.095	&	$>$4.02	$\pm$	0.89	 &	$>$0.37	&	$>$4.39	&	$>$0.165	$\pm$	0.045	&	 $>$0.015	 &	$>$0.179	\\	
070110	&	127	$\pm$	4	&	$>$25.46	&	X	&	$>$0.274	&	$>$20.61	$\pm$	 5.62	&	$>$2.57	&	$>$23.19	&	$>$0.787	$\pm$	0.274	 &	 $>$0.097	&	 $>$0.873	\\	
081203A	&	219	$\pm$	6	&	$>$3.95	&	X	&	$>$0.107	&	$>$9.71	$\pm$	2.29	 &	$>$21.14	&	$>$30.86	&	$>$0.137	$\pm$	0.032	&	 $>$0.294	&	 $>$0.429	\\	

\hline
\end{tabular}

\begin{flushleft}
$^{a}$ In units of days.\\
$^{b}$ The jet-break is identified in optical (O) or X-ray (X).\\
$^{c}$ In units of rad.\\
$^{d}$ In units of $10^{50}$ erg.\\
$^{e}$ In units of $10^{50}$ erg/s.\\
$^{f}$ No optical onset bump in GRB 090424, the initial Lorentz factor of this GRB was taken from Zou \& Piran (2010).\\
Refs: 1: Kulkarni et al.(1999); 2: Wang et al. (2015); 3: Dai et al. (2006); 4: Racusin et al. (2009);  5: Page et al. (2007); 6: Kann et al. (2007); 7: Guidorzi et al.(2011); 8: Gorbovskoy et al.(2012); 9: Zheng et al. (2012); 10: $http://www.swift.ac.uk/xrt_live_cat/00512035/$; 11: Guidorzi et al. (2014). \\
\end{flushleft}
\end{table*}

\end{document}